\documentclass[twocolumn,showpacs,preprintnumbers,amsmath,amssymb,superscriptaddress]{revtex4}
\pdfoutput=1

\usepackage{graphicx}

\usepackage{dcolumn}
\usepackage{bm}

\begin{document}

\author{William N. Plick}

\affiliation{Quantum Optics, Quantum Nanophysics, Quantum Information, University of Vienna, Boltzmanngasse 5, Vienna A-1090, Austria}
\affiliation{Institute for Quantum Optics and Quantum Information, Boltzmanngasse 3, Vienna A-1090, Austria}

\author{Mario Krenn}

\affiliation{Quantum Optics, Quantum Nanophysics, Quantum Information, University of Vienna, Boltzmanngasse 5, Vienna A-1090, Austria}
\affiliation{Institute for Quantum Optics and Quantum Information, Boltzmanngasse 3, Vienna A-1090, Austria}

\author{Robert Fickler}

\affiliation{Quantum Optics, Quantum Nanophysics, Quantum Information, University of Vienna, Boltzmanngasse 5, Vienna A-1090, Austria}
\affiliation{Institute for Quantum Optics and Quantum Information, Boltzmanngasse 3, Vienna A-1090, Austria}

\author{Sven Ramelow}

\affiliation{Quantum Optics, Quantum Nanophysics, Quantum Information, University of Vienna, Boltzmanngasse 5, Vienna A-1090, Austria}
\affiliation{Institute for Quantum Optics and Quantum Information, Boltzmanngasse 3, Vienna A-1090, Austria}

\author{Anton Zeilinger}

\affiliation{Quantum Optics, Quantum Nanophysics, Quantum Information, University of Vienna, Boltzmanngasse 5, Vienna A-1090, Austria}
\affiliation{Institute for Quantum Optics and Quantum Information, Boltzmanngasse 3, Vienna A-1090, Austria}
\affiliation{Vienna Center for Quantum Science and Technology, Faculty of Physics, University of Vienna, Boltzmanngasse 5, Vienna A-1090, Austria}

\title{Quantum orbital angular momentum of elliptically-symmetric light}

\date{\today}

\pacs{42.50.-p, 42.50.Ex, 03.67.Hk}

\begin{abstract}
\noindent We present a quantum mechanical analysis of the orbital angular momentum of a class of recently discovered elliptically-symmetric stable light fields \--- the so-called Ince-Gauss modes. We study, in a fully quantum formalism, how the orbital angular momentum of these beams varies with their ellipticity and discover several compelling features, including: non-monotonic behavior, stable beams with real continuous (non-integer) orbital angular momenta, and orthogonal modes with the same orbital angular momenta. We explore, and explain in detail, the reasons for this behavior. These features may have application to quantum key distribution, atom trapping, and quantum informatics in general \--- as the ellipticity opens up a new way of navigating the photonic Hilbert space.
\end{abstract}

\maketitle

\section{Introduction}

\noindent The study of the orbital angular momentum (OAM) of stable light modes is a newly burgeoning field. Since the seminal paper by Allen et al. \cite{A} research in this field has accelerated. Interest has been driven by the promise of access to higher dimensional Hilbert spaces (especially larger alphabet quantum key distribution \cite{QKD,QKD2}), potential probes of heretofore hidden phenomena (even astronomical events \cite{tam}), metrology \cite{jha}, use in micro-mechanics \cite{mech}, and perhaps most importantly \--- the insight the study of OAM yields into the fundamental properties of light fields themselves. Especially illuminating, is the investigation of the quantum aspects of OAM carrying light modes. Such beams have even been entangled in their orbital angular momentum degree of freedom \cite{ent}. \\

\noindent When a light field is in the paraxial regime, the total angular momentum of the classical electromagnetic field separates out into the spin and orbital components \cite{jack}. A similar decomposition exists in a fully quantum operator formalism: $\hat{\textbf{J}}=\hat{\textbf{L}}+\hat{\textbf{S}}$ for both vector and scalar fields, where $\hat{\textbf{J}}$, $\hat{\textbf{L}}$, and $\hat{\textbf{S}}$ represent the total, orbital, and spin angular momenta respectively \cite{calvo,25}. Thus these beams may be said to have a well defined OAM. The spin component (SAM) is familiar in its manifestation as the polarization of the light (right circular carrying $+\hbar$ SAM, left circular carrying $-\hbar$). This has been well known since 1909 \cite{poy} and was observed experimentally in 1936 \cite{beth}. When an SAM carrying beam is absorbed by a particle, the particle is made to spin about an axis defined by the particle's own center of mass.

\noindent The OAM of a light field is a result of the overall transverse phase structure of the beam. A particle absorbing OAM is made to rotate about the central axis of the light beam itself \--- hence the term ``orbital" (see, for example, Ref.\cite{rot}). An individual photon may carry $l\hbar$ units of OAM, where $l$ may take any integer value. All results calculated in this paper are per photon.\\

\noindent There are many classes of paraxial beams, each with their own unique properties. Two of the most familiar are the Laguerre-Gauss (LG) and the Hermite-Gauss (HG), which are the natural solutions in circular-cylindrical, and cartesian coordinates, respectively. In this paper we study the properties of a recently discovered class of OAM carrying light fields, called the Ince-Gauss (IG) beams \cite{1b,bandres}. These beams display elliptic-cylindrical symmetry. Interestingly, they become the LG modes in the limit of zero coordinate-system ellipticity, and the HG modes in the limit of infinite ellipticity. They are, in a sense, fundamental to these other light fields as they are generalizations existing in a larger mathematical space; the others being specific cases. Though all three types of beams exist in the same Hilbert space (and all three in fact span it), the IG modes are a more intuitive way to navigate through this space, as the transition between the various stable modes is controlled by a continuous real parameter which has a straightforward meaning in terms of the shape of the transverse beam profile. In a recent experiment entanglement has been generated between two IG modes \cite{us}.\\

\noindent Especially intriguing are the quantum properties of the OAM of these light fields, the study of which shall be the focus of this paper. We will find that the Ince-Gauss modes of light posses several compelling and unique properties. They offer some new insights into the nature of the orbital angular momentum of light and potentially present some new technological applications. In section two we briefly review the field of elliptical beams as a whole, with specific emphasis on those elements which are essential for the study of the orbital angular momentum of these modes. In section three we recall those parts of the established literature we require, and then proceed to the derivations and discussions which compose the main novel conclusions of this paper. We conclude this paper in section four with a brief overview and prospectus.

\section{The Ince-Gauss Modes}

\noindent For the convenience of the reader we here present a brief overview of the derivation of the Ince-Gauss modes. Further and more detailed information on the classical properties of these fields may be found in Ref.\cite{bandres}. Much of this section may be bypassed by those who have a strong familiarity with the field.\\

\noindent The Paraxial Wave Equation (PWE) describes stable beams whose transverse shape does not change under propagation. That is, those light modes for which the small angle approximation is valid. The version for scalar fields is written as

\begin{eqnarray}
\left(\nabla_{T}^{2}+2ik\frac{\partial}{\partial z}\right)\Psi(\vec{r})=0.\label{PWE}
\end{eqnarray}

\noindent Where $k$ is the wave number, $\nabla_{T}^{2}$ is the transverse Laplacian, and $\Psi(\vec{r})$ is the scalar field as a function of position. It is worthwhile to note that since we have chosen to look at scalar fields we are restricting ourself to a subset of the possible stable light modes. Vector beams (where polarization need not be constant) can exist as well and have interesting properties, see for example Ref.\cite{mon}. We wish to find the solution to this equation in the elliptic coordinate system. We can assume there is a solution of the form

\begin{eqnarray}
\mathrm{IG}(\vec{r})=E(\xi)N(\eta)e^{iZ(z)}\psi_{G}(\vec{r}).\label{IG1}
\end{eqnarray}

\noindent These are the trial Ince-Gauss solutions to the PWE. The elliptic coordinates $\xi$, $\eta$ and $z$ are related to cartesian coordinates by $x=f(z)\cosh\xi\cos\eta$, $y=f(z)\sinh\xi\sin\eta$, and $z=z$. The function $f(z)$ is the semifocal separation at position $z$. It is given by $f(z)=f(0)w(z)/w(0)$, where $w(z)$ is the beam width as a function of $z$. The point $z=0$ is defined as the point where the beam waist is at a minimum. The function $\psi_{G}$ is the fundamental Gaussian beam, it is given by

\begin{eqnarray}
\psi_{G}(\vec{r})&=&\frac{w(0)}{w(z)}\mathrm{exp}\left[\frac{-r^{2}}{w^{2}(z)}+i\frac{kr^{2}}{2R(z)}\right]\nonumber\\
& &\times\left[-i\mathrm{arctan}\left(\frac{2z}{kw^{2}(0)}\right)\right].
\end{eqnarray}

\noindent Where $R(z)$ is the radius of curvature of the phase front, and is given by $R(z)=z+k^{2}w^{4}(0)/4z$.\\

\noindent Inputting Eq.(\ref{IG1}) into Eq.(\ref{PWE}) results in three separate differential equations for $E(\zeta)$, $N(\eta)$, and $Z(z)$:

\begin{eqnarray}
 \frac{d^{2}E}{d\xi^{2}}-\epsilon\sinh(2\xi)\frac{dE}{d\xi}-[a-p\epsilon\cosh(2\xi)]E&=&0,\label{IG2a}\\
 \frac{d^{2}N}{d\eta^{2}}-\epsilon\sin(2\xi)\frac{dN}{d\eta}+[a-p\epsilon\cos(2\xi)]N&=&0,\label{IG2b}\\
 -\left(\frac{4z^{2}+k^{2}w^{4}(0)}{2kw^{2}(0)}\right)\frac{dZ}{dz}&=&p\label{IG2c}
\end{eqnarray}

 \noindent Where $a$ and $p$ are separation constants. The variable $\epsilon$ represents the ellipticity of the coordinate system and is defined as $\epsilon=2f^{2}(0)/w^{2}(0)$. Equation (\ref{IG2b}) is known as the Ince Equation - it can be transformed into Eq.(\ref{IG2a}) by making the substitution $\eta\rightarrow i\xi$. The Ince Equation was studied by Edward Lindsay Ince in 1923, who produced the eponymous solutions \cite{ince}. Equation (\ref{IG2c}) merely adds an additional phase. From now on we shall take $z=0$, for the sake of clarity. The solutions to these differential equations are known as the Ince-Gauss Beams \cite{1b,bandres}. They are given by

\begin{eqnarray}
\mathrm{IG}^{e}_{pm}(\vec{r},\epsilon)&=&\mathcal{C}C_{pm}(i\xi,\epsilon)C_{pm}(\eta,\epsilon)\mathrm{exp}\left[\frac{-r^{2}}{w^{2}(0)}\right],\nonumber\\
\mathrm{IG}^{o}_{pm}(\vec{r},\epsilon)&=&\mathcal{S}S_{pm}(i\xi,\epsilon)S_{pm}(\eta,\epsilon)\mathrm{exp}\left[\frac{-r^{2}}{w^{2}(0)}\right].
\end{eqnarray}

\noindent Where $e$ and $o$ label the even and odd modes respectively; $\mathcal{C}$ and $\mathcal{S}$ are normalization constants. The functions $C_{pm}$, and $S_{pm}$ are the even and odd Ince polynomials - which are found by first assuming they are of the form

\begin{eqnarray}
C_{2K,2n}(\eta,\epsilon)&=&\sum_{r=0}^{n}A_{r}(\epsilon)\cos(2r\eta),\label{sa} \\
C_{2K+1,2n+1}(\eta,\epsilon)&=&\sum_{r=0}^{n}A_{r}(\epsilon)\cos((2r+1)\eta), \\
S_{2K,2n}(\eta,\epsilon)&=&\sum_{r=1}^{n}B_{r}(\epsilon)\sin(2r\eta),\label{s1} \\
S_{2K+1,2n+1}(\eta,\epsilon)&=&\sum_{r=0}^{n}B_{r}(\epsilon)\sin((2r+1)\eta).\label{sb}
\end{eqnarray}

\noindent Where $K$ may take any value from zero to $n$ (except for Eq.(\ref{s1}), which may take any value between one and $n$). Substituting these expansions into Eq.(\ref{IG2b}) results in a series of recurrence relations for the weighting constants $A_{r}$ and $B_{r}$, which can be expressed as the kernel of the characteristic equation of some matrix $M$, defined via those recurrence relations. The separation constant, $a$, then takes the role of the eigenvalues. For each of the $m$ values of $a_{m}$ (where the $m$'s label the values $a$ may take in ascending order), there is an associated eigenvector of $M$ which defines the specific values of the weighting constants. \\

\noindent So, after the choice of $p$ and $m$ (which are restricted to having the same parity - that is, both even, or both odd - and $m\leq p$), the associated eigenvalue problem may be solved, giving a specific Ince polynomial for use in the Ince-Gauss equation. The restriction on the parity comes about as a matter of convention, insuring that the number of possible $m$'s \--- for a given choice of the parity of the full polynomial and the value of $p$ \--- matches the dimension of $M$. The mode numbers $p$ and $m$ are known as the order and degree, respectively. The degree of the beam is equivalent to the number of hyperbolic nodal lines (zeros in the transverse field). The order and degree together give the number of elliptic nodal lines according to the relation $(p-m)/2+\delta$, where $\delta$ is one for odd modes and zero for even modes. In the limit of zero ellipticity the IG modes become the even and odd LG beams (with an azimuthal dependence of $\cos{l\phi}$ for even, and $\sin{l\phi}$ for odd, as opposed to the typical $\mathrm{exp}(il\phi)$ dependence) with $l=m$ and $n=(p-m)/2$, where $l$ is the topological charge of the central vortex and $n$ is the radial number. A vortex is an undefined point in the transverse phase profile of a light beam, associated with a zero in the intensity pattern \--- called an optical singularity. Laguerre-Gauss with a $\mathrm{exp}(il\phi)$ azimuthal dependence contain a central vortex. In the limit of infinite ellipticity  the IG modes become the HG beams with $n_{x}=m$ and $n_{y}=p-m$ for even parity of the IG mode and, $n_{x}=m-1$ and $n_{y}=p-m+1$ for the odd modes. The root of these conversion equations is that the modes involved must all have the same Gouy phase in order to be stable, otherwise the overlap integrals are zero. \cite{1b,bandres}. \\

\noindent The IG beams are also known to be the stable resonating modes in elliptical cavities. For a more detailed analysis of the classical properties of the IG beams the interested reader is directed to Ref.\cite{bandres}.\\

\section{Properties of the Ince-Gauss Beams}

\begin{figure*}[ht]
\includegraphics[scale=0.7]{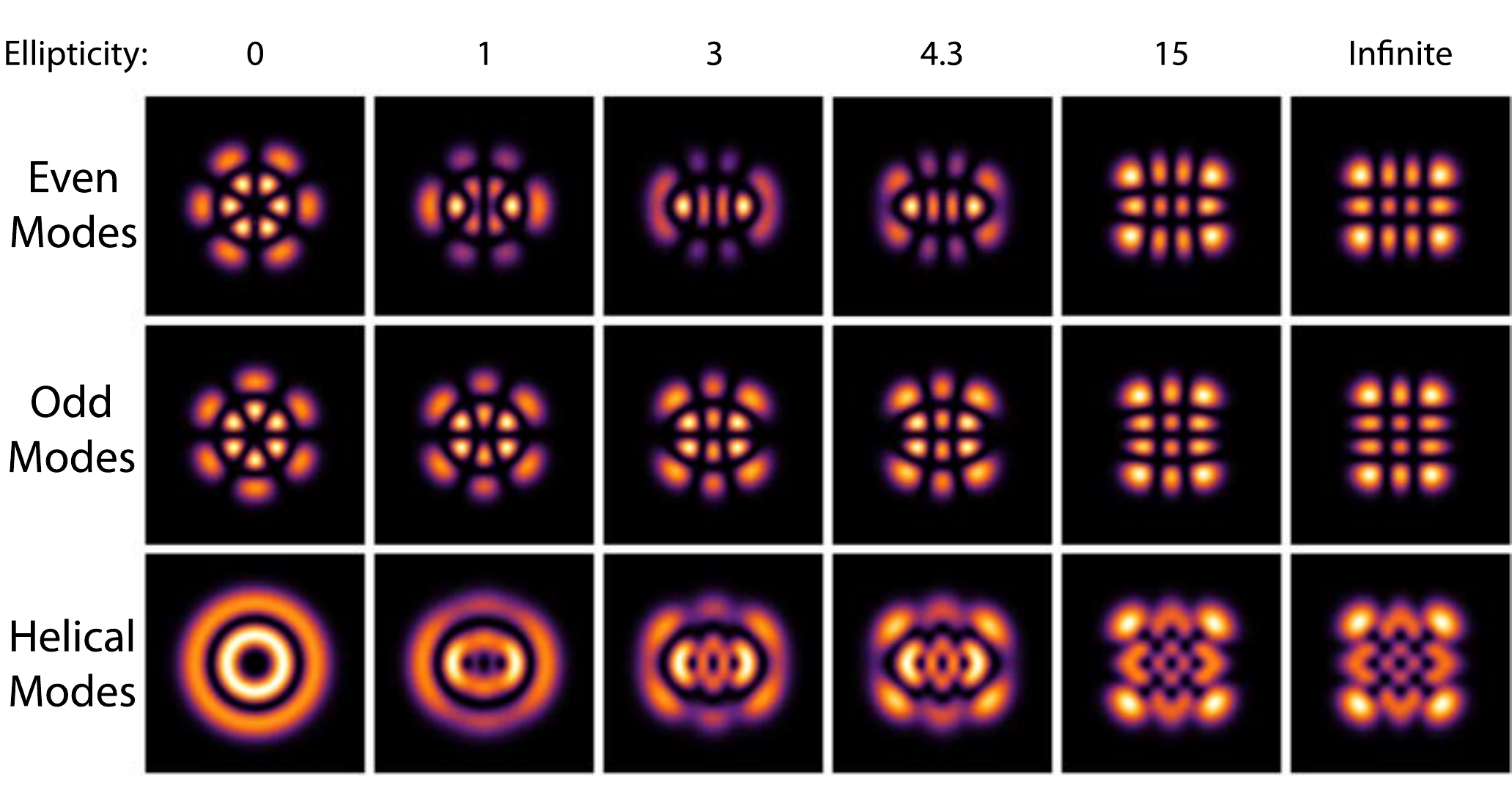}
\caption{A visualization of the transverse intensity profiles of the even, odd, and helical Ince-Gauss modes for various ellipticities and $p=5$, $m=3$. Notable is the vortex splitting and creation in the helical modes as the ellipticity increases. In the $\epsilon=0$ limit, for the helical mode, we arrive at a LG mode, and in the $\epsilon=\infty$ the IG modes become the helical hermite gauss modes. For an excellent visualization of intensity profiles side-by-side with their phase profiles, see for example Ref.\cite{bandres}.\label{pic}}
\end{figure*}

\noindent The ellipticity of the beam, specifically its relationship to the OAM, will be the primary focus of this paper. Thus it is worthwhile to briefly look at how the ellipticity affects the beam. The specifics of how the beams change as the ellipticity is varied is a function of the chosen mode numbers. However, generally speaking, as $\epsilon$ is increased the transverse intensity profile elongates along the horizontal axis. Meanwhile for Helical modes (those modes of the form $\mathrm{IG}^{e}_{pm}\pm i\mathrm{IG}^{o}_{pm}$, and which carry OAM \--- more on this shortly) the central vortex splits into a number of vortices equal to $m$, and as $\epsilon$ is increased further, new vortices may be created in the outer rings of the pattern. Interestingly, for these modes the vortices are all associated with a topological charge of one. The extremal, on axis, vortices exist at the focal points of the elliptic coordinate system defined by $\epsilon$. See Figure (\ref{pic}) for a visualization of the even and helical Ince-Gauss modes for various ellipticities and $p=5$, $m=3$. \\

\noindent While all three types of modes (even, odd, helical) have intensity zeros, only the helical modes have phase vortices. These are points where the phase of the beam is undefined and the surrounding phase undergoes a continuous change over an integer multiple of full cycles equal to the topological charge of the vortex.\\

\begin{figure*}[ht]
\includegraphics[scale=2.1]{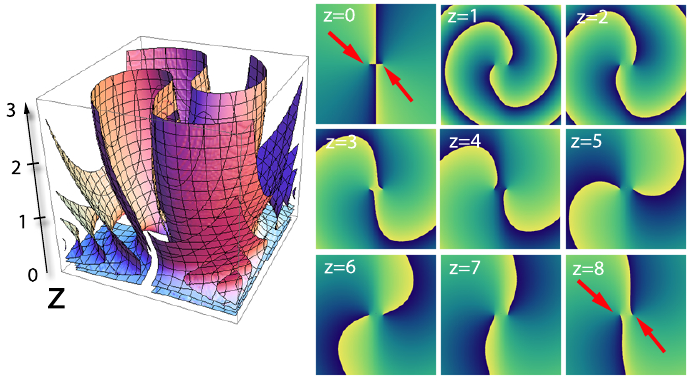}
\caption{The left-hand side depicts a Helical Ince-Gauss mode's surfaces of constant phase ($\phi=0$) as it propagates through space. The mode shown is $\mathrm{HIG}_{22}$, with $\epsilon=2$. The right-hand side depicts the transverse phase profiles at nine different points along the propagation length. The first four of these positions are also contained within the phase-surface picture. In the first and last pictures, red arrows mark the positions of the topological-charge-carrying phase singularities. These points remain stationary with respect to the foci of the elliptic coordinate system \--- although the coordinate system itself expands as the beam propagates, in the same manner as the fundamental gaussian. The equal-phase lines are seen to ``swirl" about the two phase singularities. \label{phase}}
\end{figure*}

\noindent It is also interesting to investigate the properties of the phase profile of the IG modes as they propagate through space. In Figure (\ref{phase}) we depict the equal phase surfaces, $\phi=0$, as a Helical Ince-Gauss (HIG) mode travels in the $z$-direction. The mode we depict \--- $\mathrm{HIG}_{22}$, with $\epsilon=2$ \--- contains two phase singularities which remain at the same coordinate position while the equal phase lines ``swirl" around them. We also show the full phase profile for nine different points along the propagation length. Classically, the orbital angular momentum can be seen as a result of the local phase gradient. From Fig. (\ref{phase}) it is easy to see how the complex phase profile of HIG-beams give rise to interesting OAM features. In the following section we will begin the quantum analysis.\\

\noindent The Ince-Gauss equations - like the Laguerre-Gauss equations - span the space of solutions to the PWE. As such any solution to the PWE may be decomposed into the Ince-Gauss basis. Here the ellipticity is also significant since each value it may take defines a full Hilbert space. For IG beams with the same mode numbers, the sets of orthogonal vectors which span these spaces may be transformed into each other by rotations through the Hilbert space defined by choice of ellipticity. The Ince-Gauss equations also have the familiar and useful property that they are orthonormal

\begin{eqnarray}
\int_{2D}dS\;\mathrm{IG}^{\sigma *}_{pm}\mathrm{IG}^{\sigma '}_{p'm'}=\delta_{\sigma\sigma'}\delta_{pp'}\delta_{mm'}.
\end{eqnarray}

\noindent Where $\sigma$ labels the parity and the integration takes place over the two-dimensional transverse plane. Note that IG beams with different ellipticities are typically not orthogonal - however there are some instances where they are.\\

\noindent Since the Laguerre-Gauss modes also span the space of the solutions to the PWE it is possible to decompose the Ince-Guass modes into this more familiar basis. Doing so is advantageous as the quantum Orbital Angular Momentum (OAM) properties of the Laguerre-Gauss modes are well known. The decomposition is given in Ref.\cite{1b,bandres} as

\begin{eqnarray}
\mathrm{IG}^{\sigma}_{pm}=\sum_{n,\,l}D_{nl}^{\sigma}\mathrm{LG}^{\sigma}_{nl}.\label{dec1}
\end{eqnarray}

\noindent Where the $D_{nl}$'s give the weights of the Laguerre-Gauss expansion. The required terms are given by the restriction that $p=2n+l$ (again, a requirement that comes about because the modes involved must all have the same Gouy phase to be stable), meaning that \emph{which} LG beams are needed depends only on the mode number, whereas the \emph{weights}, $D$, are dependent on the ellipticity. The parity of the Laguerre-Gauss modes must match the parity of the Ince-Gauss beam. The weights come from the overlap integrals between the IG and LG modes (which are zero if the Gouy phases differ) and are given by

\begin{eqnarray}
D_{nl}^{\sigma}&=&\mathcal{D}(-1)^{n+l+(p+m)/2}\nonumber\\
& &\times\sqrt{(1+\delta_{0l})(n+l)!n!}F_{(l+\delta_{\sigma o})/2}.\label{dec2}
\end{eqnarray}

\noindent Where $F_{(l+\delta_{\sigma o})/2}$ is the $(l+\delta_{\sigma o})/2^{\mathrm{th}}$ Fourier coefficient of of the Ince polynomial (that is, the $A_{r}$'s and $B_{r}$'s of Eq.(\ref{sa})-Eq.(\ref{sb})) associated with the beam - which is strongly $\epsilon$ dependent. The normalization constant $\mathcal{D}$ is found by requiring that $\sum_{n,\,l}D_{nl}^{2}=1$. To give a simple example, for $\mathrm{IG}_{22}^{e}$ we have the decomposition

\begin{eqnarray}
\mathrm{IG}_{22}^{e}&=&\frac{1}{\sqrt{2}\sqrt{1+\epsilon^{2}-\sqrt{1+\epsilon^{2}}}}\nonumber\\
& &\times\left(\epsilon\mathrm{LG}_{02}^{e}+(1-\sqrt{1-\epsilon^{2}})\mathrm{LG}_{10}^{e}\right).
\end{eqnarray}

\noindent For higher mode numbers the decompositions quickly become too large to report for general ellipticity. The important point to make here is that the equations are analytic.

\section{Quantum Orbital Angular Momentum of Ince-Gauss Beams}

\noindent In this section we draw upon some previously derived formalism from the field of the orbital angular momentum of light to develop a general method for analytically calculating the OAM of an Ince-Gauss photon with general mode numbers and ellipticity. The formalism is fully quantum.\\

\noindent The quantum theory of photons in Laguerre-Gauss modes is known \cite{calvo}. The creation operator for a photon in a Laguerre-Gauss Fock state may be written as

\begin{eqnarray}
\hat{a}^{\dagger}_{snl}(k_{0})=\int d^{2}\vec{q}\mathcal{LG}_{nl}(\vec{q})\hat{a}^{\dagger}_{s}(\vec{q},k_{0}).
\end{eqnarray}

\noindent Where $s$ labels the polarization mode (which may take values of $+1$ or $-1$), $\vec{q}$ is the transverse momentum vector, $\mathcal{LG}_{nl}(\vec{q})$ is the Fourier transform of the Lageurre-Gauss mode as a function of $\vec{q}$, and $k_{0}$ is the wave number of the forward-propagating plane wave. The annihilation operator is similarly defined, and they have the standard commutation relationship

\begin{eqnarray}
\left[\hat{a}_{snl}(k_{0}),\hat{a}^{\dagger}_{s'n'l'}(k_{0}')\right]=\delta_{ss'}\delta_{nn'}\delta_{ll'}\delta(k_{0}-k_{0}').
\end{eqnarray}

\noindent Since the Laguerre-Gauss modes span the space of the solutions to the PWE any paraxial one-photon state may be written as

\begin{eqnarray}
|\psi\rangle=\sum_{s,n,l}\int^{\infty}_{0}dk_{0}C_{snl}(k_{0})\hat{a}^{\dagger}_{snl}(k_{0})|0\rangle .
\end{eqnarray}

\noindent The function $C_{snl}(k_{0})$ weighs the various modes, and needs to be normalized. We can write the result of the creation operator acting on the vacuum as $|L_{nl}^{\pm}\rangle\equiv\hat{a}^{\dagger}_{nl}(k_{0})|0\rangle$, where the plus and minus indicates the sign of $l$, as a subscript in a state vector we will always write $l$ as positive.\\

\noindent It is also possible to write observables for the spin, and orbital angular momentum properties of a photon in a paraxial mode:

\begin{eqnarray}
\hat{L}_{z}=\hbar\sum_{s,n,l}l\int^{\infty}_{0}dk_{0}\hat{a}^{\dagger}_{snl}(k_{0})\hat{a}_{snl}(k_{0}),\label{lz}\\
\hat{S}_{z}=\hbar\sum_{s,n,l}s\int^{\infty}_{0}dk_{0}\hat{a}^{\dagger}_{snl}(k_{0})\hat{a}_{snl}(k_{0}).
\end{eqnarray}

\noindent The quantum Laguerre-Gauss Fock states, $|L_{nl}^{\pm}\rangle\ $, are the eigenvectors of the OAM operator $\hat{L}_{z}$, where $l$ can take any integer value.\\

\noindent Now, we have the tools necessary to study the quantum properties of the orbital angular momentum of Ince-Gauss modes of light, and begin to derive the main results of this paper. The first step is to take the Helical Ince-Gauss beams \--- that is those that are of the form $\mathrm{HIG}^{\pm}_{pm}=1/\sqrt{2}(\mathrm{IG}^{e}_{pm}\pm i\mathrm{IG}^{o}_{pm})$, and carry orbital angular momentum \--- and decompose them in terms of the Laguerre-Gauss modes. \\

\noindent Before doing this we need to write the quantum Laguerre-Gauss modes in terms of even and odd modes, as opposed to the helical modes \--- in which they are typically written

\begin{eqnarray}
|L_{nl}^{e}\rangle&=&\frac{1}{\sqrt{2}}\left(|L_{nl}^{+}\rangle+|L_{nl}^{-}\rangle\right),\\
|L_{nl}^{o}\rangle&=&\frac{1}{i\sqrt{2}}\left(|L_{nl}^{+}\rangle-|L_{nl}^{-}\rangle\right).
\end{eqnarray}

\noindent Using Eq.(\ref{lz}) we can write the action of the orbital angular momentum operator on these modes as

\begin{eqnarray}
\hat{L}_{z}|L_{nl}^{e}\rangle&=&\frac{l}{\sqrt{2}}\left(|L_{nl}^{+}\rangle-|L_{nl}^{-}\rangle\right)=i\hbar l|L_{nl}^{o}\rangle,\\
\hat{L}_{z}|L_{nl}^{o}\rangle&=&\frac{l}{i\sqrt{2}}\left(|L_{nl}^{+}\rangle+|L_{nl}^{-}\rangle\right)=-i\hbar l|L_{nl}^{e}\rangle.
\end{eqnarray}

\noindent Now, writing the quantum even (or odd) Ince-Gauss modes as superpositions of the even (or odd) Laguerre-Gauss modes in the same manner as the classical decomposition, and combining them to make the helical modes we have

\begin{eqnarray}
|I_{pm}^{\pm}\rangle=\frac{1}{\sqrt{2}}\left(\sum_{n,l}D_{nl}^{e}|L_{nl}^{e}\rangle\pm i\sum_{n',l'}D_{n'l'}^{o}|L_{n'l'}^{o}\rangle\right).
\end{eqnarray}

\noindent Where the $D$'s are the same as from Eqs.(\ref{dec1},\ref{dec2}). It should be noted that OAM is always defined relative to a given axis. We use a decomposition of LG modes with a central optical vortex, therefore the value we calculate will be with respect to the central beam axis. It is then straightforward to calculate the expectation value of the quantum OAM as a function of these expansion coefficients

\begin{eqnarray}
\langle\hat{L_{z}}\rangle=\pm\sum_{n,l}\hbar lD_{nl}^{e}D_{nl}^{o}.
\end{eqnarray}

\noindent In order to calculate these coefficients we must solve the eigenvalue problem associated with the Ince polynomials. These problems become rapidly non-trivial for increasing $p$ and $m$ - especially if we wish to allow general ellipticity. A $\mathrm{Mathematica}^{\mathrm{TM}}$ program was specifically written to perform this task. For general ellipticities the equations become far too large to report here.\\

\noindent The equations, though large, are analytically computable. A main feature that is immediately apparent is that \--- unlike the LG modes \--- the OAM is \emph{not} restricted to integer values.

\noindent Previously, fractional orbital angular momentum light beams have been studied, even using a quantum formalism \cite{frac}. However, this treatment studies fractional states created by generalizing the spiral phase patterns, used to create LG modes, to non-integer phase-step heights. These beams do not necessarily display the symmetries and resonating characteristics of stable beams (although a sub-class do). Interesting to note is the fact that this procedure requires an additional real parameter as well \--- the angle at which the phase discontinuity exists \--- perhaps parameterizing the space in analogous way to the ellipticity of the Ince-Gauss beams.\\

\noindent Though only the LG modes are eigen-modes of the orbital angular momentum operator, this is by construction. We conjecture that there is no obstacle in principle to writing an \emph{elliptic} OAM operator in a fully secondly quantized formalism, which has as its eigen-modes the IG fields. The mathematical complexity required to do this for general ellipticity however make this a non-trivial task, and it presents an interesting avenue for future work. We surmise that such an operator's eigenvalues would present as real numbers (not necessarily integers) when converted to the circular-cylindric coordinate system \--- that is, the values we here calculate. Likewise, the eigenvalues of the traditional (circular-cylindric) OAM operator would exist as non-integer average values in the elliptic basis. Independent of coordinate choice would be the type, number, and distribution of the optical vortices present in the beam. Thus, further study of the nature of optical vortices themselves seems to be promising - with the Ince-Gauss modes arising as a natural test-bed for such investigations.\\

\begin{figure}[h]
\includegraphics[scale=0.42]{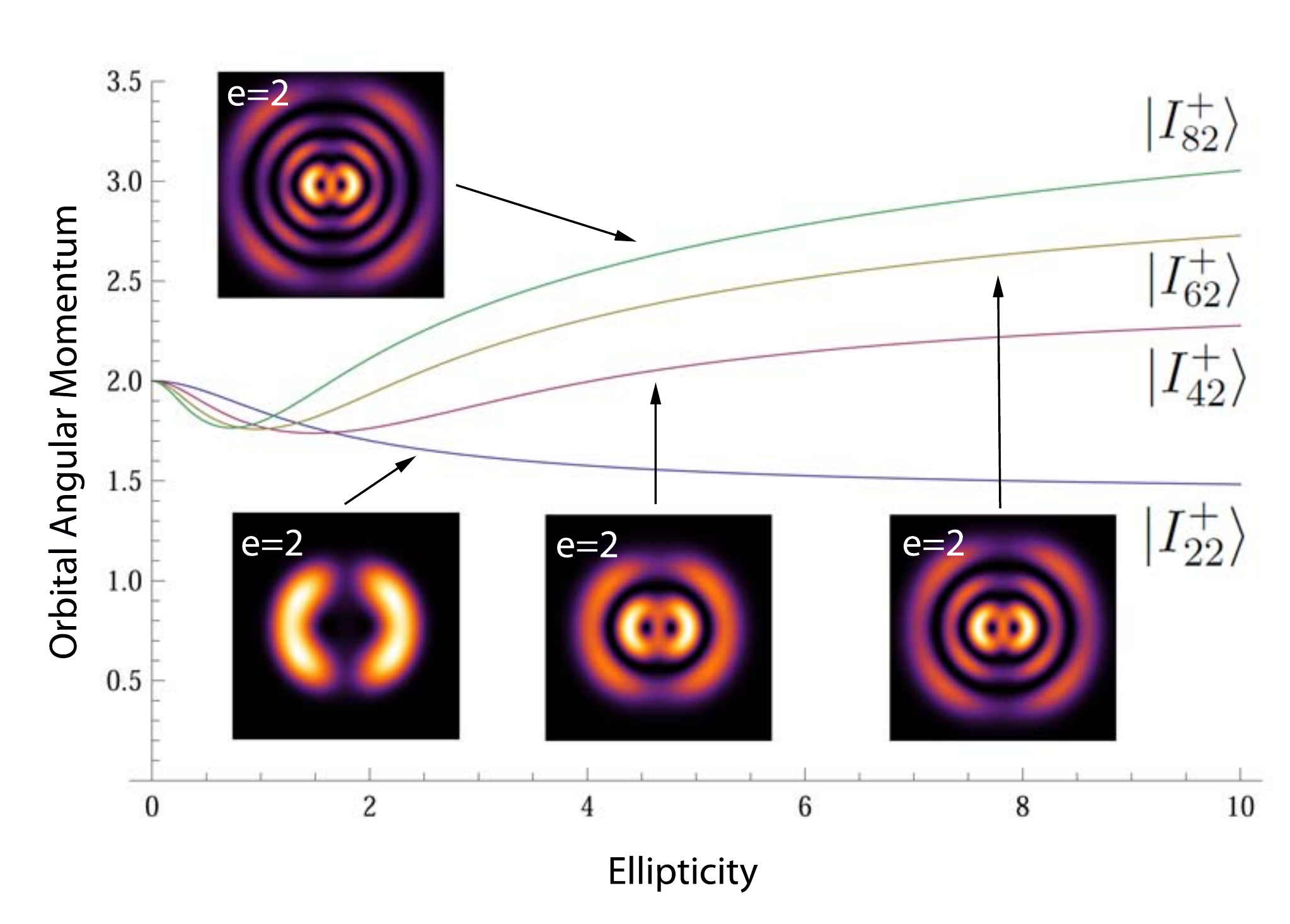}
\caption{The orbital angular momentum of four IG modes $|I^{+}_{pm}\rangle$, with the same degree (second mode number: 2) and different order (first mode number: 2, 4, 6, 8), as a function of the ellipticity of the beam. Also shown are the transverse intensity profiles of the associated beams with $\epsilon=2$. This graph highlights the real continuous (non-integer) nature of the OAM as a function of the ellipticity. And the ``turning points" of minimum OAM, a more in-depth discussion of this behavior is contained in the text.\label{ig1}}
\end{figure}

\noindent Figure (\ref{ig1}) shows the orbital angular momentum of four quantum IG modes, with the same degree number and different order numbers, as a function of the ellipticity of the beam. \\

\noindent Several interesting features are immediately apparent. First, it is clear to see the convergence on the LG modes with a topological charge of two as $\epsilon\rightarrow 0$, and the divergence into several different HG modes as $\epsilon\rightarrow\infty$. For each beam a wide spectrum of OAM values are available \textit{continuously} for stable beams. In other words the OAM may be tuned by adjusting the ellipticity.  \\

\noindent Also of note in Figure (\ref{ig1}) is the fact that the OAM does not change monotonically as the ellipticity increases. There are ``turning points" of minimum OAM. There is an intriguing semi-classical treatment of a related class of beams \--- Mathieu beams \--- where a similar phenomenon is studied \cite{mat}. Mathieu beams are generalizations to elliptical coordinates of Bessel beams (though it should be noted that both can be thought of as IG beams in the appropriate limit \cite{e}). In Ref.\cite{mat} they also find similar turning points as the ellipticity of the beam is varied and note that they appear in the proximity of a ``critical value" where new optical vortices begin to appear. This effect is seen again with ``ellipticons" which are elliptical, self-trapped beams which can exist in highly non-linear, non-local media. These objects are also described by the same equations as the IG beams \cite{elip}. A numerical, semi-classical treatment of these beams reveals turning points as well. Further research in this direction may shed more light on the relationship between OAM and optical vortices \cite{berry}. \\

\begin{figure}[h]
\includegraphics[scale=0.28]{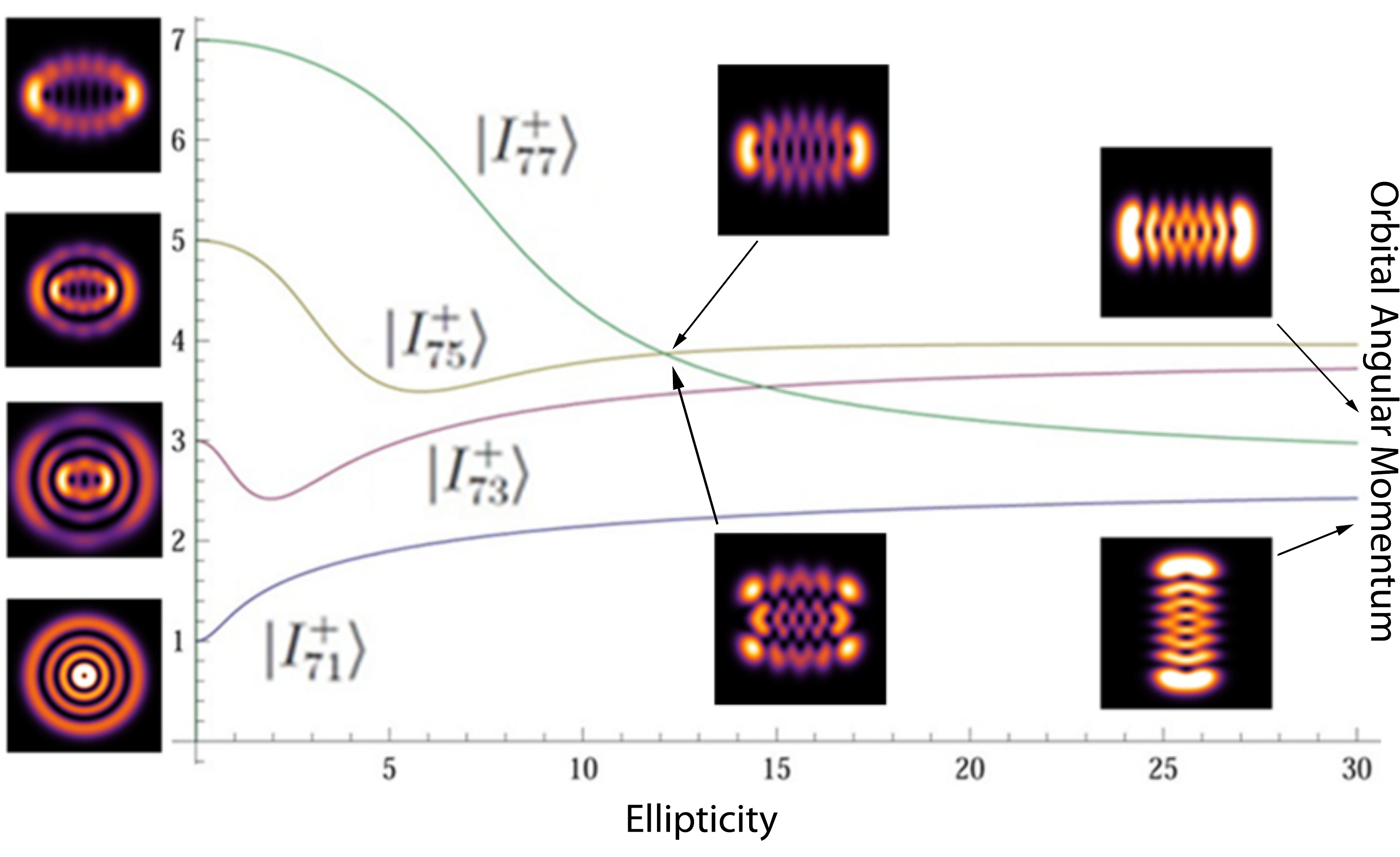}
\caption{The orbital angular momentum of four quantum IG modes, with the same order (first number (p): 7) and different degrees (second number (m): 1, 3, 5, 7), as a function of the ellipticity of the beam. To the left of the y-axis are shown the transverse intensity profiles of the associated beams with $\epsilon=2$. This graph highlights the fact that the OAM curves of different IG modes can cross. The middle insets show the transverse intensity profiles of two beams when they cross (Top: $|I^{+}_{77}\rangle$, Bottom: $|I^{+}_{75}\rangle$), with arrows pointing to the crossing point. To the right of the graph is shown the intensity profiles of two IG modes as they reach the limit of infinite ellipticity \--- becoming the same mode under a ninety degree rotation. As this occurs, both modes converge on the same OAM value. \label{ig2}}
\end{figure}

\noindent We posit, that the initial decrease in the OAM of the beam (for those beams which exhibit this behavior) is due to the topological charge carrying vortices moving apart. As they do so the torque each vortex exerts is partially counteracted by its neighbors, since in the region between any two in-line vortices the torques work against each other. This effect is most apparent in Figure (\ref{ig2}) \--- for those modes with higher degree (second mode number) the initial drop-off in OAM is more dramatic since the initial central vortex splits into more vortices along the semi-major axis of the coordinate system as the ellipticity increases. See, for example, mode $|I^{+}_{77}\rangle$, which has a steep drop off. Conversely, modes with lower degree experience less drop off. See, for example, $|I^{+}_{71}\rangle$, which has no drop-off due to the fact that there is only one on-axis vortex. \\

\noindent The subsequent increase in OAM for some modes as the ellipticity increases further is a result of the creation of off-axis vortices in the beam as the elipticity increases. The turning points can be seen when the effect of this process on the OAM begins to dominate over effect of the on-axis vortex spread. For the extremal cases \--- $|I^{+}_{77}\rangle$, and $|I^{+}_{71}\rangle$ \--- there are no turning points as one effect or the other is absent.\\

\noindent Figure (\ref{ig2}) also highlights more clearly another interesting property of the OAM of Ince-Gauss modes: The fact that the OAM curves of different IG modes can cross. This results in orthogonal modes with the same OAM at specific values of the ellipticity. This is in contrast to the LG modes where \--- due to the radial number not affecting the OAM \--- there may be an infinite number of modes with the same (integer) OAM.\\

\noindent The crossings are due to the fact that the two effects, discussed in the paragraph above, affect different modes to different extents, allowing the OAM curves to cross. For example, the $|I^{+}_{75}\rangle$ mode crosses through the $|I^{+}_{77}\rangle$ mode as the former becomes more affected by off-axis vortex creation, whereas the later is not affected by this process at all.\\

\noindent Figure (\ref{ig2}) also shows the convergence of two separate of IG modes (here, $|I^{+}_{77}\rangle$ and $|I^{+}_{71}\rangle$) on the same OAM value in the infinite limit as they become helical HG beams. The reason for this convergence can be easily demonstrated by examining the intensity distributions in the infinite limit. The beams are identical under a ninety degree rotation, thus the orbital angular momentum (being invariant under overall rotations) approaches the same value. These types of beams \--- Hermite-Gauss modes carrying OAM \--- have been studied before and dubbed the ``Helical Hermite-Gauss" modes. They have many interesting properties - for more information on them see Ref.\cite{HHG}. These two intensity diagrams also again demonstrate well the two processes (discussed previously) that affect OAM as ellipticity is varied. The $|I^{+}_{77}\rangle$ mode only experiences on-axis vortex \emph{separation}, leading to a monotonically \emph{decreasing} OAM and a horizontal line of phase vortices. By contrast the $|I^{+}_{71}\rangle$ only experiences off-axis vortex \emph{creation}, leading to a monotonically \emph{increasing} OAM and a vertical line of phase vortices.\\

\section{Interactions With Physical Systems}

\noindent Now, we address the question of how OAM transfers to systems with which the IG light-field interacts. In short, the answer is: What OAM is detected depends strongly on what composes the observing system. Consider the following cases: a.) A large, rigid physical object which interacts with the entire beam. b.) A free particle exterior to the system of vortices. c.) A free particle in close proximity to one of the vortices. d.) A device which performs a projection either in the LG or IG basis. An example would be a spatial light modulator, which is set to transform a specific mode into a Gaussian mode, which then either does or does not couple into an optical fibre \--- depending on whether the mode matches.\\

\noindent a.) The object would interact with the entire phase profile of the beam and pick up an angular momentum equivalent to the expectation value of the OAM per photon. This could also be seen as the object experiencing a torque from each singularity proportional to that singularity's topological charge. Here, it is easy to see why the OAM initially decreases as the ellipticity increases. Take for example a beam with two vortices: as the ellipticity increases and the vortices move apart their torques work against each other to an increasing degree. \\

\noindent b.) The particle would follow an elliptical orbit around the system of vortices. This has been demonstrated experimentally for the similar Mathieu-Gauss beams \cite{HMG}. \\

\noindent c.) The particle would rotate around the nearby vortex, picking up an OAM per photon of approximately the topological charge of the vortex. There would also be some second order effects from other vortices in the beam, which would be small if they were remote and potentially large if the vortices were near. \\

\noindent d.) Perhaps the most interesting case. If the system in question sorts photons according to their \emph{integer} OAM value a superposition of OAM values would occur (with weights defined as in the expansion of the IG modes in terms of LG polynomials). This may have application to quantum key distribution \--- as the ellipticity creates a larger parameter space \--- and potentially also to other tasks in quantum communication and quantum information processing. Projections into the IG basis of non-integer OAM states are also possible.\\

\noindent It is important to reemphasize that all the results reported in this manuscript are \emph{per photon}, in units of $\hbar$.\\

\section{Conclusions and Outlook}

\noindent The richness of the the Ince-Gauss modes of light present an excellent test-bed for the study of optical vortices and their connection to topological charge and the orbital angular momentum of light. This has potential application to various fields including opto-mechanics, atom trapping, and quantum informatics. Plus, insights gained from this study will continue to elucidate the nature of light itself.

\noindent In this paper we first developed a fully quantum mechanical formalism that describes the orbital angular momentum of the Ince-Gauss beams of light. We then highlighted several striking features of the OAM of these light fields as a function of the ellipticity of the the beam. We see non-monotonic behavior, where minima are reached for specific values of the ellipticity. Also, convergence on the OAM of the LG beams in the zero limit, and of the HG beams in the infinite limit is observed. Another feature is the crossing of beams with different $p$ and $m$ numbers at specific values of the OAM and ellipticity. Perhaps most importantly, we demonstrate the overall continuous nature of the OAM as the IG beams range over both integer and non-integer values in a fully quantum mechanical way for beams whose shapes are stable under propagation. That is, the OAM of a beam may be ``tuned" continuously using the IG modes. \\

\noindent It is useful to here briefly consider what use these beams may have. Given the way in which the vortices separate and move as the ellipticity increases there could be application to more advanced techniques in atomic trapping, building on an already burgeoning field which thus far makes use of simpler beam profiles only (see for example Ref.\cite{trap}). Also, the ellipticity of the beam may have use in quantum key distribution as it opens up a new continuous parameter space in which information could potentially be hidden, increasing robustness of QKD schemes to potential attacks. There may also be some application to quantum informatics in general since the Ince-Gauss light modes exist naturally as superpositions of the LG modes in a stable beam. These open possibilities all present potential paths for further research.

\section*{Acknowledgements}

\noindent This work was supported by the ERC Advanced Grant QIT4QAD, and the Austrian Science
Fund FWF within the SFB F40 (FoQuS) and W1210-2 (CoQuS).

\end{document}